\documentclass[sigconf]{acmart}
\usepackage{bookmark}
\usepackage{amsmath, flushend, graphicx, booktabs, mathtools, stfloats, subfigure, url}
\usepackage{balance}
\usepackage{geometry}
\usepackage{subfiles}
\usepackage{import}
\usepackage[utf8]{inputenc}
\usepackage{array}
\usepackage{siunitx}
\usepackage[english]{babel}
\usepackage[flushleft]{threeparttable}
\usepackage{amsmath}
\usepackage{adjustbox}
\usepackage{longtable}
\usepackage{dcolumn}
\usepackage{rotating}
\usepackage[font=scriptsize,labelfont=bf]{caption}
\usepackage{tikz}
\usepackage{mathtools}
\usepackage[font=footnotesize,labelfont=bf,
   justification=justified,
   format=plain]{caption}
\usepackage{natbib}
\newcommand\norm[1]{\left\lVert#1\right\rVert}
\newcommand{\ra}[1]{\renewcommand{\arraystretch}{#1}}
\DeclareMathOperator*{\maxi}{min}

\copyrightyear{2021}
\acmYear{2021}
\setcopyright{acmlicensed}
\acmConference[SAC '21]{The 36th ACM/SIGAPP Symposium on Applied Computing}{March 22--26, 2021}{Virtual Event, Republic of Korea}
\acmBooktitle{The 36th ACM/SIGAPP Symposium on Applied Computing (SAC '21), March 22--26, 2021, Virtual Event, Republic of Korea}
\acmPrice{15.00}
\acmDOI{10.1145/3412841.3442065}
\acmISBN{978-1-4503-8104-8/21/03}

\begin{document}

\title[Utilizing Textual Reviews in Latent Factor Models for Recomm. Systems]{Utilizing Textual Reviews in Latent Factor Models for Recommender Systems}

\author{Tatev Karen Aslanyan}
\affiliation{%
  \institution{Erasmus University Rotterdam}
  \streetaddress{P.O. Box 1738}
  \city{Rotterdam}
  \country{the Netherlands}
  \postcode{NL-3000 DR}
}
\email{tatevkaren@gmail.com}

\author{Flavius Frasincar}
\orcid{0000-0002-8031-758X}
\affiliation{%
  \institution{Erasmus University Rotterdam}
  \streetaddress{P.O. Box 1738}
  \city{Rotterdam}
  \country{the Netherlands}
  \postcode{NL-3000 DR}
}
\email{frasincar@ese.eur.nl}

\begin{abstract}
Most of the existing recommender systems are based only on the rating data, and they ignore other sources of information that might increase the quality of recommendations, such as textual reviews, or user and item characteristics.  Moreover, the majority of those systems are applicable only on small datasets (with thousands of observations) and are unable to handle large datasets (with millions of observations).
We propose a recommender algorithm that
combines a rating modelling technique (i.e., Latent Factor Model) with a topic modelling method based on textual reviews (i.e., Latent Dirichlet Allocation), and we extend the algorithm such that it allows adding extra user- and item-specific information to the system.
We evaluate the performance of the algorithm using Amazon.com datasets with different sizes, corresponding to 23 product categories. After comparing the built model to four other models we found that combining textual reviews with ratings leads to better recommendations. Moreover, we found that adding extra user and item features to the model increases its prediction accuracy, which is especially true for medium and large datasets.
\end{abstract}




\maketitle

\section{Introduction}
\label{section: introduction}


Throughout the last decade, the importance of the Web as a medium for business and electronic transactions has increased drastically, forcing the IT to rapidly develop as well, making humans daily life much easier and more efficient. On its turn, this large development in IT has increased the popularity of online shopping and services. Making purchases online instead of buying products from physical shops, which can be very time-consuming, is one of the major consequences of IT development. However, this large increase in online sales has not only led to an increase in the number of customers but also an increase in the number of products and variety of these products. Therefore, when making purchase decisions, users are forced to process large amounts of information. According to \citep{LeeLee2004}
this information overload has a big impact on the human decision process and quality. Hence, it affects people's online purchase experience significantly. Therefore, to overcome this problem, one usually rely on suggestions from others, who have more experience on the topic \cite{Senecal2010}. This idea is used in the recommender systems aiming to employ various sources of information to recommend products to the users by inferring their interests. Besides solving the problem of information overload, the use of recommender systems also results in increased sales, customer satisfaction and loyalty \citep{Schafer1999}, which explains increasing popularity of these systems. On the one hand, the information overload motivates the use of recommender systems in order to make the users' online purchases more convenient. On the other hand, the increasing variety of ways that users can discover, evaluate, and review online products motivates companies and researchers to create even more revealing recommender algorithms, which will enable to sell more products.

The Web enables users to provide their personal feedback about the product that they have purchased in the form of ratings and textual reviews. Assuming that the past interests and preferences are often good proxies of future choices, the previous interactions between items and users can be used for predicting which items might be interesting for a user in the future. Therefore, in order to correctly recommend the users their desired products, one should predict how the user will respond to a new product~\citep{Aggarwal2016}. Recommender systems are usually categorized as: Collaborative Filtering systems based on ratings data \citep{Sarwar2001}, Content-Based systems based on content (often textual) data \citep{Lops2011}, and Hybrid systems that combine these two types of systems \citep{Cano2019}. Most of the existing recommender systems are of the first type (based only on ratings), and they ignore the enormous information incorporated in the users' review texts \citep{Zhang2006}. Ignoring such an important source of information, that can potentially increase the accuracy of recommendations, seems not optimal. Moreover, adding extra user- and item-specific information, not included in the ratings or textual reviews, to the recommender system might also increase the quality of its recommendations \citep{Buono2001, Zhen2009, Kawasaki2017}.

\begin{figure}
\centering
\includegraphics[width = 9cm,height = 5.6cm]{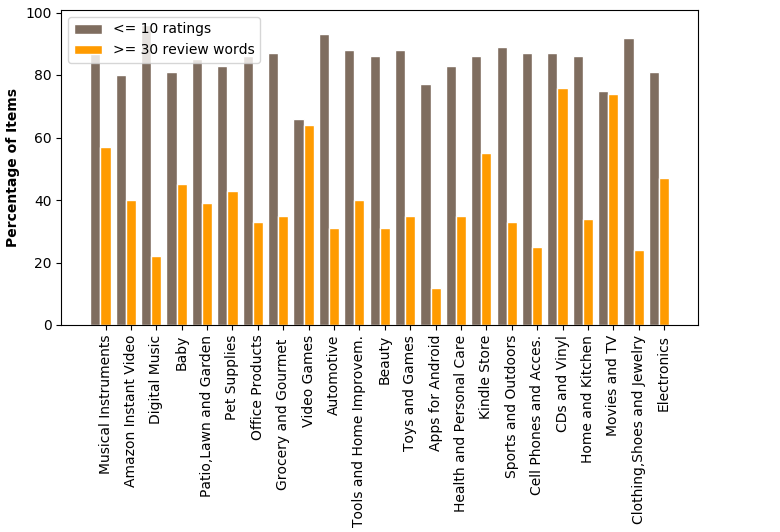}
\caption[Percentage of ratings and reviews per item]{Percentage of ratings and reviews per item. Brown (dark grey for black and white print) bars represent the percentage of items with less than 10 ratings per category. Green (light grey for black and white print) bars represent the percentage of items having on average more than 30 words per product category.}
\label{fig:1}
\end{figure}

Figure \ref{fig:1} presents the percentage of items having less than 10 ratings and more than 30 words in their review text per product category in the datasets of the largest e-commerce Amazaon.com \citep{McAuley2015}. We observe that for almost all product categories it holds that at least 80$\%$ of items have very few ratings (less than 10) while over 40$\%$ of items have long textual reviews (with more than 30 words per review). Therefore, textual reviews can be considered as a potential source of information that can used to complement the absent ratings to increase the prediction accuracy of the recommender system.
In this paper, we propose a recommender algorithm that is based on product ratings as well as textual reviews of customers, and it allows adding extra user- and item-specific information in order to make product recommendations. We focus on Latent Factor Models (also called Matrix Factorization) for modeling the item features and user preferences in a shared topic space and Topic Models (more precisely Latent Dirichlet Allocation) for modelling review features.

\section{Related Work}
\label{section: related work}

Although, there exist a large amount of literature regarding recommender systems that are based on a single type of data, such as ratings or textual reviews, there have been only few attempts of combining user-item ratings and textual reviews to uncover the latent rating and latent review dimensions \citep{Toscher2010, Blei2011, Leskovec2013b, Ling2014, Bao2014}.  \citep{Toscher2010} combined the predictions of the Latent Factor Model (LFM) with the predictions of the neighborhood model to generate more accurate recommendations. A similar approach was taken in case of the recommender system of `Bellkor's Pragmatic Chaos', the Netflix Prize contest winner \citep{Koren2009b}. This system compares the watching and searching habits of similar users, and then recommends movies that share the characteristics with movies that are highly rated by the current user. Since then, LFM became the most popular Collaborative Filtering (CF) technique used for both rating and item recommendations \citep{Rossetti2013}.

\cite{Blei2011} have developed an algorithm called Collaborative Topic Modeling, combining CF and probabilistic topic modeling, which recommends scientific papers to an online community of users. Authors found that the proposed recommender system, based on both contents of articles and users' ratings, performs better than the recommender system based on standard Matrix Factorization (MF) methods. Among all the Hybrid recommender systems, one of the most known systems combining ratings with textual reviews for making recommendations is the Hidden Factors and Topics (HFT) algorithm proposed in \cite{Leskovec2013b}. HFT combines latent rating dimensions learned by LFM, with latent review topics learned by the topic modeling technique Latent Dirichlet Allocation (LDA), in order to make rating predictions. \cite{Leskovec2013b} stated that, the HFT algorithm results in highly interpretative textual labels for the hidden rating dimensions helping to `justify' ratings with review text, and in increased prediction accuracy of the recommender system. Another example of a recommender algorithm that combines ratings with textual reviews has been introduced in \cite{Ling2014}, called Ratings Meet Reviews (RMR). The proposed method is a probabilistic generative model combining the topic modeling technique LDA with the MF method for ratings. The main difference between HFT and RMR is the way the authors combine the two models. More specifically, HFT uses the MF method to model the ratings, whereas RMR uses a mixture of Gaussian distributions. \cite{Ling2014} found that RMR outperforms the standard MF based approach and results in similar prediction accuracy compared to HFT. The TopicMF algorithm introduced by \cite{Bao2014} is also an example of a recommender algorithm combining ratings and reviews in order to make recommendations for the users. TopicMF uses biased MF for modeling the ratings and uses Non-negative Matrix Factorization (NMF) for modeling the latent topics in the textual reviews. The main difference between this algorithm and the earlier mentioned recommender algorithms is that it uses NMF instead of LDA as the topic modeling approach. The final example related to the model introduced in this study is the Rating-Boosted Latent Topics (RBLT) algorithm introduced by \cite{Tan2016}. RBLT used LDA for extracting topics from the reviews like HFT and RMR and it also uses MF for modeling the ratings like HFT. The main difference between RBLT and HFT is that HFT uses item-features in rating prediction and topic-distributions as a regularization for these item-features, whereas the RBLT includes the topic-distributions in the rating prediction procedure but not in the regularization term. \cite{Tan2016} found that adding textual reviews to the CF system increases its prediction accuracy significantly.

One similarity that is shared by the previously surveyed papers is that they propose to use textual reviews as well as ratings to model item features and user preferences in a shared topic space and consequently bring them into LFM to generate recommendations. Our research will also be focused on utilizing recommender systems with the MF approach by using product ratings as well as textual reviews of customers.

There have been also few attempts of building a recommender system that allows adding user- or item-specific characteristics, not present in the rating or review data \citep{Buono2001}, \citep{Kawasaki2017}, \citep{Zhen2009}. \cite{Buono2001} and \cite{Kawasaki2017} introduced CF recommender systems that also allow adding user- and item-specific features on the top of the ratings. As extra user and item information \cite{Kawasaki2017} used the browsing data.  The CF system extension in \cite{Kawasaki2017} has been done by adding extra rows and columns to the user-item rating matrix. However, all these extended recommenders that allow adding user or item features to the system, are based only on ratings.
To our knowledge, there are no studies of recommender systems combining ratings and textual reviews that also allow adding extra user or item information to the system. Another limitation of the existing literature is that most of the proposed recommender systems are modeled and implemented on a dataset consisting of very few product categories or a small number of observations.

To address the previously identified limitations we propose a recommender algorithm called LDA-LFM, which combines the topic-modeling technique LDA with the rating-modeling method LFM and allows adding (latent) extra user- and item-specific features to make recommendations. LDA-LFM is a generalization of the HFT model proposed by \cite{Leskovec2013b}, but it will use an alternative approach for model regularization and will allow adding extra user- and item-specific features to the recommender system. These extra features will behave as additional factors in MF driving the ratings following the approach proposed in \cite{Kawasaki2017}, while these extra features do not appear in the topic modelling method LDA.
This system is applied on both small and large datasets,
with or without a large number of product categories.
\section{Rating and Review Models}
\label{section: methodology}

In this section, we introduce all models and techniques used to build and evaluate the proposed LDA-LFM model. We describe the technical details and optimization approach of LFM and the topic modeling technique LDA used in this study.

\subsection{Latent Factor Model for Recommendations}
\label{subsection: Latent Factor Model}
In CF recommender systems, the Latent Factor Model (LFM), also called Matrix Factorization, has become very popular especially after the earlier mentioned Netflix Prize Contest \citep{Koren2009a, Koren2010a}.
Usually, the rating matrix contains lots of missing elements, thus suffers from the sparsity problem. In order to overcome this problem, LFM uses the idea of dimensionality reduction to estimate and fill in all missing entries of the sparse user-item rating matrix.
The goal of dimensionality reduction is to rotate the axis system such that the pairwise correlations between dimensions can be removed and a large sparse matrix can be decomposed into smaller and dense matrices. Accordingly, the reduced, rotated, and complete data matrix representation can be efficiently estimated from a sparse data matrix. The key idea of the Matrix Factorization method is that any m x n sparse matrix R with rank k $<$ min\{m,n\} can be approximated by rank-k matrices in the following way \citep{Strang2009}:
\begin{equation}
R \approx PQ^T
\end{equation}
where P and Q are m x k and n x k matrices, respectively. So, the user-item sparse matrix R is approximately equal to the product of P and Q matrices, such that the vectors of R can be represented by the rows of matrix P and columns of matrix Q. Stated differently, in LFM, sparse ratings matrix R is decomposed into the product of two low-rank rectangular matrices P, the user matrix, and Q, the item matrix, where both P and Q have the same rank k.
Each row of matrix P and each column of matrix Q are referred as $\textit{latent factors}$. Let us define by p$_u$ the $\textit{u}$th row of user matrix P, the user factor representing the affinity of user u towards the rating matrix R, and by q$_i$ the $\textit{i}$th row of item matrix Q, the item factor representing the affinity of $\textit{i}$th item towards the rating matrix R.
Since, some users have a tendency to give higher ratings while other users are more prone to provide lower ratings, and that some products have a tendency to be highly rated compared to other products, baseline predictions (biases) should also be taken into account. \citep{KorenBell2011} referred to biases as the observed variation in rating values due to the effects associated with either items or users independent of any interaction. Correspondingly, the estimate of each rating of the $\textit{u}$th user about $\textit{i}$th item, denoted by r$_{ui}$, can be expressed as follows:
\begin{equation}
\begin{aligned}
&\hat{r}_{ui} = \alpha + b_i + b_u + q_i^Tp_u
\label{eq:1}
\end{aligned}
\end{equation}
where $\alpha$ represents the global average of all ratings (an offset parameter), b$_u$ and b$_i$ represent the user and item biases, respectively. Accordingly, the error which arises in this estimation is defined as e$_{ui}$ = r$_{ui}$- $\hat{r}_{ui}$ and in order to learn the latent factors p$_u$ and q$_i$ the following optimization problem should be solved, where we minimize the regularized squared error \citep{Koren2009b}:
\begin{equation}
\begin{aligned}
&\text{arg} \
\maxi_{\hat{\Theta}} \dfrac{1}{\mid{    \mathcal{T}}\mid} \sum_{u,i\in    \mathcal{T}}(e_{ui})^2 + \lambda\Omega(\Theta)\\
& \text{arg} \
\maxi_{\hat{\Theta}} \dfrac{1}{\mid{    \mathcal{T}}\mid} \sum_{u,i\in    \mathcal{T}}(r_{ui}-\hat{r}_{ui})^2 + \lambda\Omega(\Theta)\\
&\text{arg} \
\maxi_{\hat{\Theta}} \dfrac{1}{\mid{\mathcal{T}}\mid} \sum_{u,i\in    \mathcal{T}}(r_{ui}-(\alpha + b_i + b_u + q_i^Tp_u))^2 + \lambda\Omega(\Theta)\\
&\Omega(\Theta) = \norm{q_i}_2^2 + \norm{p_u}_2^2 + \norm{b_i}_2^2 + \norm{b_u}_2^2
\label{eq:2}
\end{aligned}
\end{equation}
$\mathcal{T}$ represents the corpus of all ratings (in the training set), $\Theta$ = \{$\alpha$, b$_u$, b$_i$, p$_u$, q$_i$\} is the parameter space of the model. The objective function in Equation \ref{eq:2} can be seen as quadratic loss function which quantifies the loss of accuracy when an element of rating matrix R is approximated by low-rank factorization. $\norm{ . }_2^2$ represents the squared Frobenius norm, also called the L$_2$ norm. We use regularization to prevent model overfitting, which is required especially when the dataset used for fitting the model contains a large number of features, like it is in our case.
The constant $\lambda$ from Equation \ref{eq:2}, which is often referred as the regularization constant, determines the level of regularization and controls how hard unnecessary features in the model are penalized. Determining the value of $\lambda$ is a trade-off between prediction variance and bias. One popular way of determining the optimal value is Grid Search.

One of the most popular ways of solving the optimization problem defined in Equation \ref{eq:2} is Stochastic Gradient Descent (SGD) \citep{Zhang2004, KorenBell2011, Leskovec2013b, Zhao2016}. Other typical methods which can be considered as possible alternatives to the SGD method, like the Limited-memory Broyden-Fletcher-Goldfarb-Shanno (L-BFGS) \citep{Zhu1997} or Orthant-Wise Limited-memory Quasi-Newton (OWL-QN) \citep{GalenJianfeng2007}, work very slowly when the model is fitted on a large training dataset and performing it by one machine is sometimes intractable. SGD addresses these issues because it scales well with both big data and with the size of the model, therefore it is preferred in this analysis. However, even though the method itself is simple and fast, it is known as a ``bad optimizer" because it is prone to finding local optimum instead of a global optimum.
A popular technique designed to improve the performance of SGD method is the Adaptive Moment Estimation (Adam) introduced by \cite{Kingma2015}. Adam is the extended version of the SGD (with momentum). The main difference compared to the SGD (with momentum), which uses a single learning rate for all parameter updates, is that Adam algorithm defines different learning rates for different parameters. The algorithm calculates the individual adaptive learning rates for each parameter based on the estimates of the first two moments of the gradients. 

\subsection{Latent Dirichlet Allocation}
\label{subsection: Latent Dirichlet Allocation}
Each text review provided by a user, represented as a bag of words, contains valuable information, which can potentially increase the prediction accuracy of the recommender system. For this reason, the textual reviews should be modeled and analyzed.
Latent Dirichlet Allocation (LDA) introduced by \cite{Blei2003} is one of the most popular text mining methods in the context of recommender systems.
Therefore, we will use LDA as a topic modeling technique in this analysis in order to uncover the hidden dimensions in the user review texts.

There are three main entities defined in this method: words, documents, and corpora. The entity $\textit{word}$ is defined as a basic unit of a discrete data from a vocabulary, $w_{d,j}$ where j $\in$ \{1,2,...,N$_d$\}, which indicates the index of the word in document d. These words are represented in the form of a vector where the $\textit{j}$th element of this vector takes value 1 and remaining all elements take value 0. The entity $\textit{document}$ is a sequence of N words denoted by $d\in\mathcal{T}$ such that N$_d$ represents the number of words in document d. Finally, the entity $\textit{corpus}$ is defined as a collection of documents denoted by $\mathcal{T}$ = (d$_1$,d$_2$,...,d$_M$), where M is the number of all documents in the corpus. For simplicity we will use indices to identify documents.

LDA makes a few important assumptions regarding the model. Firstly, it assumes that words carry strong semantic information and that documents discussing similar topics will use similar words. Therefore, latent topics are discovered by identifying a bag of words in a corpus that frequently occur together in a document. Secondly, LDA assumes that documents are probability distributions of latent topics and topics are probability distributions of words. So, every document consists of a certain amount of topics and each of these topics is a distribution of words. Therefore, the model assumes that there are in total K latent topics. Then, LDA assigns to each document d a K-dimensional topic distribution $\theta_d$ drawn from a Dirichlet distribution represented in the form of a stochastic vector, such that the $\textit{k}$th entry of it, $\theta_{d,k}$, represents the fraction of words in document d which discuss $\textit{k}$th topic. Stated differently, the likelihood that words in document d will be about topic k is equal to $\theta_{d,k}$. Furthermore, each topic k is a distribution of words represented by $\phi_k$ such that each word has a particular likelihood of being used in the topic k.\\
Let us denote by z$_{d,j}$ the topic assigned to the $\textit{j}$th word in document d. Then the LDA model is defined as follows:
\begin{itemize}
    \item $\theta_d$ $\sim$ $\mathcal{DIR}$($\gamma$) \ \    with d $\in$ \{1,...,M\}
    \item $\phi_k$ $\sim$ $\mathcal{DIR}$($\nu$) \ \   with k $\in$ \{1,...,K\}
    \item z$_{d,j}$ $\sim$  $\textit{Multinomial}(\theta_d)$
    \item w$_{d,j}$ $\sim$  $\textit{Multinomial}(\phi_{z_{d,j}})$
\end{itemize}
where $\gamma$ represents the parameter of the Dirichlet distribution for document-topic distribution $\theta_d$ and $\nu$ is the parameter of the Dirichlet distribution for word-topic distribution $\phi_k$. \ z$_{d,j}$ represents the topic assigned to the \textit{j}th word in document d. We assume that the total number of words in vocabulary is V.  Moreover, we denote the likelihood function of z$_{d,j}$ conditional on topic mixture of document d, $\theta_d$, p(z$_{d,j}\mid\theta_d$) as follows:
\begin{equation}
\begin{aligned}
&p(z_{d,j}\mid\theta_d) = \theta_{d,z_{d,j}}
\label{eq:3}
\end{aligned}
\end{equation}
Consequently, the probability of $\textit{j}$th word in document d, w$_{d,j}$, conditional on the chosen topic z$_{j}$ denoted by $p(w_{d,j}\mid z_{d,j},\nu)$ is defined as follows:
\begin{equation}
\begin{aligned}
p(w_{d,j}\mid z_{d,j},\nu)= \phi_{z_{d,j},w_{d,j}}
\label{eq:4}
\end{aligned}
\end{equation}
Furthermore, using the definition of the Dirichlet probability distribution, the conditional topic distribution is defined as follows:\begin{equation}
\begin{aligned}
&p(\theta\mid\gamma) = \dfrac{\Gamma(\sum_{d=1}^D\gamma_d)}{\prod_{d = 1}^D\Gamma(\gamma_d)}\theta_1^{\gamma_1-1}\dots\theta_d^{\gamma_d-1}
\label{eq:5}
\end{aligned}
\end{equation}
where $\theta_d$ $>$ 0 and $\Gamma(\cdot)$ represents the Gamma function. Consequently, the joint distribution of a topic $\theta$, K topics z, and N words w is defined as follows:
\begin{equation}
\begin{aligned}
&p(\theta, z, w\mid\gamma,\nu) = p(\theta\mid\gamma)\prod_{j=1}^{N}p(z_{j}\mid\theta)p(w_{j}\mid z_{j},\nu)
\label{eq:6}
\end{aligned}
\end{equation}
where N = $\sum_{d = 1}^M$ N$_d$.
Using the properties of discrete and continuous random variables' distributions, the marginal distribution of document d is defined as follows:
\begin{equation}
\begin{aligned}
&p(w\mid\gamma,\nu) = \int p(\theta\mid\gamma)\prod_{j=1}^N\sum_{z_{j}}p(z_j\mid\theta_d)p(w_{j}\mid z_{j},\nu)d\theta_d
\label{eq:7}
\end{aligned}
\end{equation}
Consequently, using Equations \ref{eq:3}, \ref{eq:4} and \ref{eq:7} the likelihood of a text corpus $\mathcal{T}$ conditional on the word distribution $\phi$, topic distribution $\theta_d$ and topic assignments z is defined as follows:
\begin{equation}
p(\mathcal{T}\mid \gamma,\nu,z) =  \prod_{d\in\mathcal{T}}\bigg(\int p(\theta_d\mid\gamma)\prod_{j=1}^{N_d}\sum_{z_{d,j}}\theta_{d,z_{d,j}}\phi_{z_{d,j},w_{d,j}}\bigg)d\theta_d
\label{eq:8}
\end{equation}
This expression can also be rewritten in terms of the topic distribution $\theta_d$ and word distribution $\phi$, in the following way:
\begin{equation}
\begin{aligned}
&p(\mathcal{T}\mid \theta,\phi,z) =  \prod_{d\in\mathcal{T}} \prod_{j=1}^{N_d}\theta_{d,z_{d,j}}\phi_{z_{d,j},w_{d,j}}
\label{eq:9}
\end{aligned}
\end{equation}
where parameters $\theta$ and $\phi$ should be estimated, which we denote by $\Phi$, such that $\Phi$ = \{$\theta, \phi$\}. Then, the log-transformation of the conditional corpus probability p($\mathcal{T}$$\mid$ $\theta$, $\phi$, z) is defined as follows:
\begin{equation}
\begin{aligned}
&l(\mathcal{T}\mid\theta, \phi, z) =\sum_{d\in\mathcal{T}} \sum_{j=1}^{N_d}log\big(\theta_{d,z_{d,j}}\phi_{z_{d,j},w_{d,j}}\big)
\label{eq:10}
\end{aligned}
\end{equation}
Typically, in order to estimate the LDA model parameters, Variational Bayesian (VB) methods or sampling approaches based on Markov Chain Monte Carlo (MCMC) sampling are being used \citep{Blei2003, Griffiths2004}.

\begin{figure}[h]
\centering
\includegraphics[width = 8.3cm,height = 3.6cm]{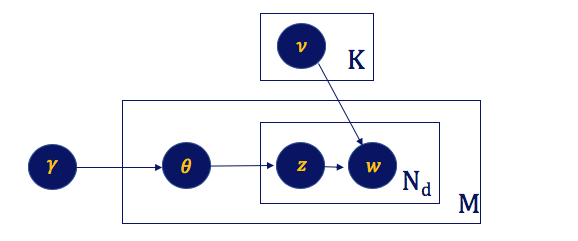}
\caption[LDA Visualization]{LDA Visualization. K is the number of topics; M is the number of documents; N$_d$ is the length of document d; $\theta_d$ represents the topic distribution of document d which follows the Dirichlet distribution with parameter $\gamma$; $\nu$ is the corresponding parameter of the word probability distribution of the topic k; w$_{d,j}$ is a word in document d and z$_{d,j}$ is that word's topic.}
\label{figSGD}
\end{figure}

Figure \ref{figSGD} visualizes the dependencies among the LDA model parameters. High $\gamma$ indicates that it is likely that each document contains a mixture of most of the topics. Conversely, low $\gamma$ indicates that each document contains only few of the topics. Furthermore, high $\nu$ indicates that each topic contains most of the words of that topic, whereas small $\nu$ means that each topic contains only small amount of words. The parameters $\gamma$ and $\nu$ are at the $\textit{corpus level}$ which are both assumed to be sampled once in the process of corpus generation. The random variable $\theta_d$ is the only variable at the $\textit{document level}$, sampled once per document. Finally, the variables z$_{d,j}$ and w$_{d,j}$ are at the \textit{word level} sampled once for each word per document.

\subsection{LDA-LFM Model}\label{LDA-LFM Model}
The model that we design, called Latent Dirichlet Allocation-Latent Factor Model (LDA-LFM), aims to combine two main core ideas of two methods discussed in Sections \ref{subsection: Latent Factor Model} and \ref{subsection: Latent Dirichlet Allocation} to to uncover both hidden dimensions in ratings and textual reviews, respectively.

As it was mentioned earlier, one of the three entities on which topic modeling is based on, is the \textit{document} entity. Therefore, the concept of `document' in the LDA-LFM model should be defined properly. There are different ways of defining this concept which should be based on the textual reviews. One can simply consider each text review of user u and item i as a document, denoted by d$_{ui}$. On the other hand, one can define a document as a set of all reviews corresponding to item i, denoted by d$_i$. Finally, the set of all reviews provided by a user u as a document, denoted by d$_u$. \cite{Leskovec2013b} found that the second definition, where the concept of a document is defined as the set of all reviews of item i (d$_i$), leads to the best model performance. The motivation behind this choice is that when users provide feedback about the products in terms of textual reviews, they discuss more often the characteristics of the product rather than discussing their personal preferences. Therefore, we will define the concept of documents in LDA-LFM in the similar way as in \cite{Leskovec2013b}.

The idea behind the LDA-LFM model is to find the K-dimensional topic distribution $\theta_i$ of each item using textual reviews of item i which shows the extent to which each topic k is discussed across all the reviews for item i. Consequently,
these topic distributions are used as item-factors in combination with user-factors in the Latent Factor Model to fully predict all user-item ratings. In Section \ref{subsection: Latent Factor Model} we stated that parameter q$_i$ is the rating factor possessing the properties of item i that can be reviewed by users, whereas in Section \ref{subsection: Latent Dirichlet Allocation} we stated that parameter $\theta_i$ is the topic distribution of words that appear in those reviews. Assuming that, if an item i has a certain property, then it will correspond to a particular topic discussed in that item's textual review, such that q$_{i,k}$ and $\theta_{i,k}$ are positively correlated, we need to define the exact relation between these two parameters. However, q$_{i,k}$ and $\theta_{i,k}$ cannot be considered as being equal since the topic distribution $\theta_i$ is a stochastic vector describing topic probabilities while latent item factor q$_i$ can take an arbitrary value in $\bold{R}^K$. Stating that q$_i$ is a stochastic vector like $\theta_i$ would result in a loss of power in the proposed model and changing the structure of the topic distribution $\theta_i$ to make it more similar to q$_i$ will lead to the loss of probabilistic power in the model. In order to not encounter these problems, the transformation of q$_i$ to $\theta_i$ should satisfy monotonicity, q$_i$ $\in$ $\bold{R}^K$, and $\sum_k \theta_{i,k}$ = 1 assumptions.
The following transformation satisfies all these criteria:
\begin{equation}
\begin{aligned}
&\theta_{i,k} = \dfrac{exp(\kappa q_{i,k})}{\sum_{k' = 1}^K exp(\kappa q_{i,k'})}
\label{eq:431}
\end{aligned}
\end{equation}
where the parameter $\kappa$ controls for the reaching of the highest possible value of the transformation, often called `pickiness' parameter. Large value of $\kappa$ indicates that users discuss only the most important topic, whereas small $\kappa$ indicates that users discuss all topics equally. We define the transformation, in such a way that, when $\kappa$ $\rightarrow$ $\infty$, $\theta_i$ $\rightarrow$ $\iota$ (unit vector with 1 for the largest value of $q_{i,k}$), and, when $\kappa$ $\rightarrow$ 0, $\theta_i$ converges to a uniform distribution. To make sure that the word distribution for topic k ($\phi_{k}$) is a stochastic vector, the following transformation of $\phi_{k}$ is defined with an introduction of a new variable $\psi$:
\begin{equation}
\begin{aligned}
&\phi_{k,w} = \dfrac{exp(\psi_{k,w})}{\sum_{w'}exp(\psi_{k,w'})}
\label{eq:20}
\end{aligned}
\end{equation}
where $\psi_k$ $\in$ R$^V$ is used as a natural parameter for the topic distribution $\phi_k$ $\in$ R$^V$, where V is the size of the vocabulary. Correspondingly, it holds that $\sum_{n}$$\phi_{k,n}$ = 1. Then the objective function of the LDA-LFM model is defined as follows:\\
\begin{equation}
\begin{aligned}
&f(\mathcal{T}\mid\Theta, \Phi, \kappa, z) = \sum_{{u,i}\in\mathcal{T}} (r_{u,i}-\hat{r}_{u,i})^2 + \lambda(\norm{p_u}_2^2 + \norm{b_i}_2^2 + \norm{b_u}_2^2)\\
& - \mu l(\mathcal{T}\mid\theta, \phi, z)
\label{eq:13}
\end{aligned}
\end{equation}
where $\Theta$ = \{$\alpha$, b$_u$, b$_i$, p$_u$, q$_i$\} and $\Phi$ = \{$\theta, \phi$\} represent the set of parameters of the LFM and LDA model, respectively. The first term of Equation \ref{eq:13} represents the prediction error corresponding to LFM, the second term represents the regularization of model parameters b$_u$, b$_i$, p$_u$ and the third term represents the log-likelihood of the corpus of ratings and users from Equation $\ref{eq:10}$. The parameter $\mu$ $\in$ $R^+$ trades-off the importance of these two effects. We observe that in the LDA-LFM model, the regularization of q$_i$ is different compared to the standard Matrix Factorization case, the regularization term does not contain the norm of q$_i$. More specifically, the third term of Equation \ref{eq:13} behaves as a regularization for q$_i$ \citep{Leskovec2013b}.

\subsection{LDA-LFM with Extra Features}\label{extendedmodel}
As it was mentioned earlier, the proposed recommender system should allow adding extra user- and item-specific features. A key aspect in adding extra features to the system is to better describe users and items, in order to better predict the preferences of those users for different items. Examples of user features are user demographics such as age, living area, gender, occupation, etc. \citep{Grivolla2014}. If our goal is to build a movie recommender, then the genre, year of its release, name of the director, can all be interpreted as item characteristics, which can be added to system for making better recommendations.
\cite{Buono2001}, \cite{Kawasaki2017} and \cite{Zhen2009} introduced CF recommender systems that allow adding user- and item-specific features on the top of ratings. We will follow the approach of \cite{Buono2001} and \cite{Kawasaki2017}, who proposed adding extra rows and columns to the user-item rating matrix representing the extra features added to LFM.
\begin{figure}[h]
\centering
\includegraphics[width = 8.5cm,height = 4cm]{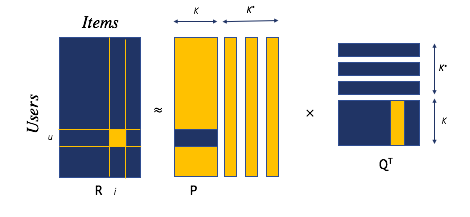}
\caption[Matrix Factorization with Extra Features]{Matrix Factorization of User-Item Rating Matrix with Extra Features. Matrix R is the rating matrix. Matrices P and Q are the user factor and item factor matrices, respectively. To both matrices P and Q are added by K* = 3 extra user-columns and item-rows. These extra K$^*$ columns/rows do not appear in the LDA model while the first K ones do.}
\label{fig:3}
\end{figure}
Figure \ref{fig:3} visualizes an example of the Matrix Factorization model extended with three extra features. The main idea is to add the same amount of both extra user- and item-specific features. This assumption is necessary because LFM, which is used as rating modelling technique in the LDA-LFM model, requires matrix multiplication of two matrices with dimensions NUsers x K and K x NItems. This matrix multiplication is only possible when the number of columns in user-factor matrix P is equal to the number of rows of item-factor matrix Q. The extra features denoted by K$^*$ from Figure \ref{fig:3} do not appear in the LDA model and represent non-review factors that affect the review ratings.
\subsection{LDA-LFM Model Fitting}
Our goal is to find the solution to the optimization problem of Equation \ref{eq:13}, which is:
\begin{equation}
\begin{aligned}
&\text{arg} \
\maxi_{\Theta, \Phi, \kappa, z} f(\mathcal{T}\mid\Theta, \Phi, \kappa, z)
\label{eq:14}
\end{aligned}
\end{equation}
where the corpus $\mathcal{T}$ is given. The LDA-LFM model defines the following uterative stochastic optimization procedure of two steps:
\begin{equation}
\begin{aligned}
&\textbf{for i in Niter}\\
&\ \ \text{Solve} \ \ \text{arg} \
\maxi_{\Theta, \Phi, \kappa} f(\mathcal{T}\mid\Theta, \Phi, \kappa, z^{(t-1)}) \rightarrow \text{Update} \ \ \Theta^{(t)}, \Phi^{(t)}, \kappa^{(t)}\\
&\ \ \text{Sample} \ \ z_{d,j}^{(t)} \text{with} \ \ p(z_{d,j}^{(t)} = k) = \theta_{d,k}\phi_{k,w_{d,j}}^{(t)}\\
&\textbf{end for}
\label{eq:15}
\end{aligned}
\end{equation}
where Niter is the number of iterations, d $\equiv$ d$_{u,i}$ represents the review or set of reviews (document) of item i by user u.

In the first step of this optimization procedure from Equation \ref{eq:15} we fix the topic assignments for each word, i.e., the value of latent variable z and we solve the objective function with respect to $\Theta$, $\Phi$ and $\kappa$.
We use the Adam Optimizer for learning the rating related model parameters $\Theta$ = \{$\alpha$, b$_u$, b$_i$, p$_u$, q$_i$\}, but also the review related parameters $\Phi$ = \{$\theta$, $\phi$ \}, and $\kappa$. As it was mentioned earlier, $\theta$ $\in$ $\Phi$ and q $\in$ $\Theta$ are linked through Equation \ref{eq:431}. So, we do not use the textual reviews in order to fit the document-topic distribution $\theta$ using the LDA approach. Instead, we determine $\theta$ using q. Since, we introduced a transformation of $\phi$, to ensure that it is a stochastic vector, instead of learning $\phi$ we learn the parameter $\psi$. Once we learn the $\psi$, by using the transformation defined in Equation \ref{eq:20}, the topic-word distribution $\phi$ can be determined.  Moreover, using the same optimization approach, we also learn the parameter $\kappa$.

In the second step of this iterative procedure,
using the updated parameter values $\Phi$ = \{$\theta$, $\phi$ \} determined in the first step by the Adam Optimization, we randomly assign a topic k to each word, with a probability that is proportional to the likelihood of the occurrence of that topic with that particular word \citep{Blei2011}. That is, the topic assignment probability of assigning \textit{k}th topic to a word w$_{u,i,j}$ for user u, item i and in \textit{j}th position p(z$_{w_{u,i,j}}$ = k) is proportional to the product of topic probability for user u, item i ($\theta_{u,i,k}$), and word probability used for that topic ($\phi_{k,w_{u,i,j}}$). We assume that the terms z$_{w_{u,i,j}}$ and z$_{u,i,j}$ are equivalent (z$_{w_{u,i,j}}$ $\equiv$ z$_{u,i,j}$).
We iterate through all documents and word positions, d and j, respectively, in order to update the corresponding topics assigned to those terms. Finally, we repeat these two steps for Niter times and report the prediction accuracy of the model corresponding to the last iteration.
\section{Evaluation}
\label{section: evaluation}
As a prediction accuracy measure we use the Mean Squared Error (MSE) determined as follows:
\begin{equation}
MSE = \dfrac{\sum_{(u,i) \in \mathcal{T}^*}^{} (\hat{r}_{u,i}-r_{u,i})^2}{\mid\mathcal{T}^*\mid}
\end{equation}
where $\mathcal{T}^*$ represents the corpus of all ratings in the test set, $r_{u,i}$ represents the real rating from the test data for user u and item i, and $\hat{r}_{u,i}$ is the corresponding predicted rating. MSE can take only non-negative values. Moreover, a lower value of MSE is an indication of better performing model.

It is worth to mention that the analysis is performed on a commodity machine with a Core i7 processor, 2.2 GHz frequency, and 252gb memory space using the programming language Python 3.7.
\subsection{Data}
\label{subsection: data}
In this research, we use a collection of datasets provided corresponding to the 23 product categories supplied by one of the largest e-commerce company in the world, Amazon.com. This data without duplicates was prepared by Julian McAuley. It consists of 142.8 million product reviews and a metadata for 9.4 million products, spanning a period of 18 years, from May 1996 to July 2014 \cite{McAuley2015, McAuleyHe2016}. The chosen dataset is of 5-core type, that is, the data set excludes all customers and products having less than 5 reviews. The review dataset includes feedbacks of Amazon customers in the form of ratings, textual reviews, and helpfulness score. Meanwhile, the metadata includes various characteristics of the product: price, brand, descriptions, category information, image features and links of `also viewed' and `also bought' products. The raw review data, after removing duplicates and excluding users or items with less than 5 reviews, consists of 42.13 million reviews.

Table \ref{desc1} presents the general overview of the datasets of all product categories. We observe that all datasets are highly sparse and contain a very large amount of missing ratings. For almost all datasets it holds that the average star rating is approximately equal to 4. Moreover, the average number of words per review is at least 18 and at most 67. Finally, the smallest dataset, \textit{Musical Instruments}, consists of 0.5 million reviews and the largest dataset, \textit{Electronics}, consists of approximately 8 million reviews.
\begin{table*}[ht]
\ra{1.2}
\footnotesize\centering
\begin{tabular}{|p{11cm}|p{6cm}|p{6cm}|p{7cm}|p{6cm}|p{6cm}|p{6cm}|}
\hline
\multicolumn{1}{l}{\textbf{Dataset}} &\multicolumn{1}{l}{\textbf{NUsers}}&\multicolumn{1}{l}{\textbf{NItems}}&\multicolumn{1}{l}{\textbf{NReviews}}&\multicolumn{1}{l}{\textbf{A. W.}}&\multicolumn{1}{l}{\textbf{A. R.}}&\multicolumn{1}{l}{\textbf{Sparsity}}\\
\hline
\multicolumn{1}{l}{\textbf{Electronics}} &\multicolumn{1}{l}{4,201,696}&\multicolumn{1}{l}{476,002}&\multicolumn{1}{l}{7,824,482}&\multicolumn{1}{l}{43}&\multicolumn{1}{l}{4.012}&\multicolumn{1}{l}{0.00039}\\
\multicolumn{1}{l}{\textbf{Clothing, Shoes and Jewelry}}&\multicolumn{1}{l}{3,117,268}&\multicolumn{1}{l}{1,136,004}&\multicolumn{1}{l}{5,748,920}&\multicolumn{1}{l}{26}&\multicolumn{1}{l}{4.145}&\multicolumn{1}{l}{0.00016}\\
\multicolumn{1}{l}{\textbf{Movies and TV}} &\multicolumn{1}{l}{2,088,620}&\multicolumn{1}{l}{200,941}&\multicolumn{1}{l}{4,607,047}&\multicolumn{1}{l}{58}&\multicolumn{1}{l}{4.187}&\multicolumn{1}{l}{0.00110}\\
\multicolumn{1}{l}{\textbf{Home and Kitchen}} &\multicolumn{1}{l}{2,511,610}&\multicolumn{1}{l}{410,243}&\multicolumn{1}{l}{4,253,926}&\multicolumn{1}{l}{36}&\multicolumn{1}{l}{4.099}&\multicolumn{1}{l}{0.00041}\\
\multicolumn{1}{l}{\textbf{CDs and Vinyl}} &\multicolumn{1}{l}{1,578,597}&\multicolumn{1}{l}{486,360}&\multicolumn{1}{l}{3,749,004}&\multicolumn{1}{l}{67}&\multicolumn{1}{l}{4.403}&\multicolumn{1}{l}{0.00049}\\
\multicolumn{1}{l}{\textbf{Cell Phones and Accessories}}&\multicolumn{1}{l}{2,261,045}&\multicolumn{1}{l}{319,678}&\multicolumn{1}{l}{3,447,249}&\multicolumn{1}{l}{30}&\multicolumn{1}{l}{3.811}&\multicolumn{1}{l}{0.00048}\\
\multicolumn{1}{l}{\textbf{Sports and Outdoors}} &\multicolumn{1}{l}{3,117,268}&\multicolumn{1}{l}{1,136,004}&\multicolumn{1}{l}{3,268,695}&\multicolumn{1}{l}{34}&\multicolumn{1}{l}{4.145} &\multicolumn{1}{l}{0.00034}\\
\multicolumn{1}{l}{\textbf{Kindle Store}} &\multicolumn{1}{l}{1,406,890}&\multicolumn{1}{l}{430,530}&\multicolumn{1}{l}{3,205,467}&\multicolumn{1}{l}{42}&\multicolumn{1}{l}{4.232}&\multicolumn{1}{l}{0.00050}\\
\multicolumn{1}{l}{\textbf{Health and Personal Care}} &\multicolumn{1}{l}{1,851,132}&\multicolumn{1}{l}{252,331}&\multicolumn{1}{l}{2,982,326}&\multicolumn{1}{l}{33}&\multicolumn{1}{l}{4.110}&\multicolumn{1}{l}{0.00063}\\
\multicolumn{1}{l}{\textbf{Apps for Android}} &\multicolumn{1}{l}{1,323,884}&\multicolumn{1}{l}{61,275}&\multicolumn{1}{l}{2,638,173}&\multicolumn{1}{l}{18}&\multicolumn{1}{l}{3.996}&\multicolumn{1}{l}{0.00325}\\
\multicolumn{1}{l}{\textbf{Toys and Games}} &\multicolumn{1}{l}{1,342,911}&\multicolumn{1}{l}{327,698}&\multicolumn{1}{l}{2,252,771}&\multicolumn{1}{l}{33}&\multicolumn{1}{l}{4.150
}&\multicolumn{1}{l}{0.00051}\\
\multicolumn{1}{l}{\textbf{Beauty}} &\multicolumn{1}{l}{1,210,271}&\multicolumn{1}{l}{249,274}&\multicolumn{1}{l}{2,023,070}&\multicolumn{1}{l}{31}&\multicolumn{1}{l}{4.149
}&\multicolumn{1}{l}{0.00067}\\
\multicolumn{1}{l}{\textbf{Tools and Home improvement}} &\multicolumn{1}{l}{1,212,468}&\multicolumn{1}{l}{260,659}&\multicolumn{1}{l}{1,926,047}&\multicolumn{1}{l}{36}&\multicolumn{1}{l}{4.130}&\multicolumn{1}{l}{0.00061}\\
\multicolumn{1}{l}{\textbf{Automotive}} &\multicolumn{1}{l}{851,418}&\multicolumn{1}{l}{320,112}&\multicolumn{1}{l}{1,373,768}&\multicolumn{1}{l}{30
}&\multicolumn{1}{l}{4.185}&\multicolumn{1}{l}{0.00051}\\
\multicolumn{1}{l}{\textbf{Video Games}} &\multicolumn{1}{l}{826,767}&\multicolumn{1}{l}{50,210}&\multicolumn{1}{l}{1,324,753}&\multicolumn{1}{l}{58}&\multicolumn{1}{l}{3.979}&\multicolumn{1}{l}{0.00051}\\
\multicolumn{1}{l}{\textbf{Grocery and Gourmet Food}} &\multicolumn{1}{l}{768,438}&\multicolumn{1}{l}{166,049}&\multicolumn{1}{l}{1,297,156}&\multicolumn{1}{l}{31}&\multicolumn{1}{l}{4.255}&\multicolumn{1}{l}{0.00102}\\
\multicolumn{1}{l}{\textbf{Office Products}} &\multicolumn{1}{l}{909,314}&\multicolumn{1}{l}{130,006}&\multicolumn{1}{l}{1,243,186}&\multicolumn{1}{l}{36
}&\multicolumn{1}{l}{3.979
}&\multicolumn{1}{l}{0.00105}\\
\multicolumn{1}{l}{\textbf{Pet Supplies}} &\multicolumn{1}{l}{740,985}&\multicolumn{1}{l}{103,288}&\multicolumn{1}{l}{1,235,316}&\multicolumn{1}{l}{37}&\multicolumn{1}{l}{4.111}&\multicolumn{1}{l}{0.00161}\\
\multicolumn{1}{l}{\textbf{Patio, Lawn and Garden}} &\multicolumn{1}{l}{714,791}&\multicolumn{1}{l}{105,984}&\multicolumn{1}{l}{993,490}&\multicolumn{1}{l}{37}&\multicolumn{1}{l}{4.006} &\multicolumn{1}{l}{0.00130}\\
\multicolumn{1}{l}{\textbf{Baby}} &\multicolumn{1}{l}{531,890}&\multicolumn{1}{l}{64,426}&\multicolumn{1}{l}{915,446}&\multicolumn{1}{l}{41}&\multicolumn{1}{l}{4.118} &\multicolumn{1}{l}{0.00270
}\\
\multicolumn{1}{l}{\textbf{Digital Music}} &\multicolumn{1}{l}{478,235}&\multicolumn{1}{l}{266,414}&\multicolumn{1}{l}{836,006}&\multicolumn{1}{l}{41
}&\multicolumn{1}{l}{4.540}&\multicolumn{1}{l}{0.00066}\\
\multicolumn{1}{l}{\textbf{Amazon Instant Video}} &\multicolumn{1}{l}{426,922}&\multicolumn{1}{l}{23,965}&\multicolumn{1}{l}{583,933}&\multicolumn{1}{l}{28}&\multicolumn{1}{l}{4.316}&\multicolumn{1}{l}{0.00571}\\
\multicolumn{1}{l}{\textbf{Musical Instruments}} &\multicolumn{1}{l}{339,231}&\multicolumn{1}{l}{83,046}&\multicolumn{1}{l}{500,176}&\multicolumn{1}{l}{45
}&\multicolumn{1}{l}{4.244
}&\multicolumn{1}{l}{0.00178}\\
\hline
\end{tabular}
\caption[Overview of datasets for all product categories]{Overview of datasets. The following statistics per dataset are reported: number of users (NUsers), number of items (NItems), number of reviews (NReviews), average number of words in textual reviews
after removing the stopwords (A. W.), average star rating (A. R.), and sparsity of the user-item rating matrix (Sparsity).}
\label{desc1}
\end{table*}
\subsection{Data Preparation}
\label{subsection: Data Separation}
In order to correctly evaluate the chosen model, we split the data into three datasets: training, validation, and test sets. We fit the model on the training data and find a set of optimal model parameters (hyperparmeter tuning) using the validation set. Finally, we use the test set for predicting the ratings and calculating model accuracy measures using the optimal set of parameters from the hyperparmeter tuning. For data separation we use the common 80/20 splitting rule. In order to have enough observations to correctly fit the model, we put 80$\%$ of all observations in the training set, while the remaining 20$\%$ we equally divided into the test and validation sets. However, splitting the data into training, test and validation sets, when some of the users and items appear only in the test set and not in the training set, will result in a loss of information about those users and items during the training of the model. Therefore, after randomly splitting the data into train and test set, we make sure that there is no user or item that is present in the test set but not present in the training set by removing these.

For implementing the topic modeling technique LDA, the review data should be cleaned. Therefore, we perform a few Natural Language Processing (NLP) tasks on the textual reviews in review tuples by using the \textit{Natural Language
Tool Kit (NLTK)} library of the programming language Python.
Firstly, we apply tokenization to all review texts, which are provided as a group of sentences, and transform them into a group of words.  Secondly, we transform them to lower case words and remove from these tokenized reviews the common English stop words and one-letter words. Subsequently, all special characters, digits, punctuation and single or multiple spaces are removed. Next, we apply lemmatization to the processed review text, for removing inflectional endings and holding the dictionary (base) form of a word only, known as the lemma of the word.
Finally, we combine all those cleaned  reviews corresponding to the same item and create a corpus of documents, where each document contains all reviews (represented in the form of a group of words) corresponding to one item.

\subsection{Parameter Initialization}\label{ParamInit}
Different methods introduced earlier contain various parameters which should be initialized. We initialize the offset $\alpha$ by averaging over all ratings in the training set. Vectors b$_i$, b$_u$ and matrices P, Q are initialized using the random normal distribution. The fitting procedure of all models have been performed by the Adam Optimization with the learning rate 0.01. As initial value for $\kappa$ we take the value 1, which will be updated by the Adam Optimization while fitting the model. For each model we run 35 iterations based on \citep{Bao2014} (with 20 iterations), \citep{Koren2010a} (with 20-35 iterations), \citep{Jin2018} (with 30 iterations) while updating model parameters in $\Theta$, $\Phi$, and $\kappa$ in each iteration. The prediction accuracy of the model is reported based on the last model corresponding to \textit{35}th iteration, assuming that the last model, after all the updates, is the best performing model. As a common practice, for the LDA model we set both parameters $\gamma$ and $\nu$ equal to 0.1. Following the approach of \cite{Leskovec2013b} we perform the analysis with the number of latent factors in the LFM model (K) and number of topics in the LDA model (L) equal to 5.

LDA-LFM contains two regularization parameters, $\lambda$ and $\mu$. We tune this set of two parameters using the Grid-Search method, which fits the model for every specified combination of these two parameters and evaluates each of these models using validation set. As a result, the most accurate model specification, per product category, is then used in the main model prediction applied to the test dataset.
Following the approach of \cite{Leskovec2013b}, for $\lambda$ we use values \{0, 0.001, 0.01, 1, 10\} as a possible values in the Grid-Search, while for regularization constant $\mu$ we use the values \{1, 10, 100, 1000, 10,000\}. As it was mentioned in Section \ref{LDA-LFM Model}, we set the number of documents in the LDA model equal to the number of items in the data, where each document represents all reviews of an item in the training set. We set the size of the vocabulary equal to 5000, by keeping most 5000 frequent words from the corpus of all documents built from the item reviews present in training set.
\subsection{Baseline Models}
\label{subsection: Baseline Models}
In order to test for the performance of the proposed LDA-LFM model, we use four other models based on different algorithms. Then we compare the prediction accuracy of the LDA-LFM model with the performance of the following models:
\\\\
\noindent\textbf{\textit{Offset Model}:} the predicted rating for all users is the same and is equal to  the global average $\alpha$.
\\\\
\noindent\textbf{\textit{Baseline Rating Model}:} the predicted rating $\hat{r}_{ui}$ = $\alpha$ + $\bar{r}_u$ + $\bar{r}_i$ with $\alpha$ representing the global average rating, $\bar{r}_u$ the average difference between user ratings and the global average $\alpha$, $\bar{r}_i$ represents the average difference between item ratings and the global average $\alpha$.
\\\\
\noindent\noindent\textbf{\textit{LFM}:} standard Latent Factor Model model corresponding to Equation \ref{eq:1}.
\\\\
\noindent\textbf{\textit{LDAFirst}:} in this model the user feedback in the form of ratings will be used as an input for the standard LFM, while the textual reviews will be used as an input for the LDA model described in Section \ref{subsection: Latent Dirichlet Allocation}. The key difference between this method and the proposed method LDA-LFM is that, in LDAFirst the topic-distributions $\theta_i$ are sampled from a Dirichlet distribution, where each document is treated as the set of all reviews corresponding to item i, and they are used to set the q$_i$ values, which stays constant while modeling the ratings. Thus we do not learn the q$_i$ parameter of LFM during the iterative optimization procedure and we only update the parameters b$_i$, b$_u$ and p$_u$ using the Adam Optimization. In the LDA-LFM model, we do not use the LDA method for determining the topic-distributions $\theta_i$. After that we sample the word topics, we learn $\Phi$ = \{$\theta$, $\phi$\} using Adam Optimization as in Equation \ref{eq:15} (where q$_i$ is dependent on $\theta_i$ by means of Equation \ref{eq:431}). Since we start this method by the LDA model and use its output ($\theta_i$) as an input for the LFM method ($q_i$), we refer to this method as LDAFirst.

\section{Results}
\label{section: results}
Firstly, we perform the analysis for the case when no extra features are added to the LDA-LFM model, assuming that the number of topics in the LDA model is equal to the number of latent factors in the LFM model with K $\in$ \{5,10\} \cite{Leskovec2013b}. Correspondingly, in order to analyse the impact of adding extra latent features (user- and item characteristics) on the performance of recommender system based on the proposed LDA-LFM model, such that the number of topics in LDA has value 5 and the number of latent factors with  4 different values of extra features K$^*$ $\in$ \{1, 2, 3, 4\}.

Table \ref{ress1} presents the prediction per product category with the number of topics equal to 5. We observe that for the majority of supplied datasets it holds that the Offset and Baseline models perform the worst, with large MSE values, compared to the LFM, LDAFirst, and LDA-LFM models. Only for \textit{Video Games} and \textit{Tools and Home Improvements} datasets the Offset method performs better than the LFM and LDAFirst models. Moreover, we observe that compared to the Offset and Baseline models, standard LFM improves the recommender systems prediction accuracy for almost all datasets, except datasets \textit{Patio, Lawn and Garden}, \textit{Video Games}, and \textit{Tools and Home Improvement}. This can be seen by the large difference between the MSE values corresponding to LFM, and MSE values corresponding to the Offset and Baseline models.
We also observe that the MSE's corresponding to the LFM and LDAFirst models are very close to each other for the majority of datasets. This means that the LDAFirst model does not improve the prediction accuracy a lot compared to the standard LFM model. LDAFirst slightly outperforms LFM in case of the datasets \textit{Baby},\textit{Office Products} \textit{Grocery and Gourmet Food}, \textit{Apps for Android}, \textit{CDs and Vinyl}. From Table \ref{ress1} we observe that the LDA-LFM model outperforms all other models in almost all datasets, with its lowest MSE values. Last two columns of Table \ref{ress1} present the percentage decrease in the MSE of the LDA-LFM model compared to the LFM and the LDAFirst models, respectively.
Both improvement columns consist mostly of positive entries. We observe that the proposed LDA-LFM results in at least 0.24$\%$ (\textit{Health and Personal Care}) and at most 14.12$\%$ (\textit{Kindle Store}) improvement in prediction accuracy, compared to the standard LFM. From Imp.$_{[5]/[4]}$ we observe that the proposed LDA-LFM results in at least 0.28$\%$ (\textit{Electronics}) and at most 14.12$\%$ (\textit{Kindle Store}) improvement, compared to the LDAFirst model. The improvement columns in Table \ref{ress1} contain also few negative values which correspond solely to small datasets (\textit{Musical Instruments}, \textit{Amazon Instant Video}, \textit{Patio, Lawn and Garden}).

Table \ref{ress1} also reports the average MSE over all datasets per model in case of K = 10. We observe that average MSE's per model with K = 5 and K = 10 are similar.

\begin{table*}[ht]
\ra{1.16}
\footnotesize\centering
\begin{tabular}{|p{5cm}|p{2.5cm}|p{2.5cm}|p{2.5cm}|p{2.5cm}|p{2.5cm}|p{2.5cm}|p{2.5cm}|}
\hline
\multicolumn{1}{l}{\textbf{Dataset}} &\multicolumn{1}{l}{\textbf{Offset$_{[1]}$}}&\multicolumn{1}{l}{\textbf{Base.$_{[2]}$}}&\multicolumn{1}{l}{\textbf{ LFM$_{[3]}$}}&\multicolumn{1}{l}{\textbf{LDAFirst$_{[4]}$}}&\multicolumn{1}{l}{\textbf{LDA-LFM$_{[5]}$}}&\multicolumn{1}{l}{\textbf{Imp.$_{[5]/[3]}$}}&\multicolumn{1}{l}{\textbf{Imp.$_{[5]/[4]}$}}\\
\hline
\multicolumn{1}{l}{\textbf{Electronics}} &\multicolumn{1}{l}{2.909}&\multicolumn{1}{l}{2.345}&\multicolumn{1}{l}{1.789}&\multicolumn{1}{l}{1.789}&\multicolumn{1}{l}{1,780}&\multicolumn{1}{l}{0.28$\%$}&\multicolumn{1}{l}{0.28$\%$}\\
\multicolumn{1}{l}{\textbf{Clothing, Shoes and Jewel.}} &\multicolumn{1}{l}{2.275}&\multicolumn{1}{l}{2.122}&\multicolumn{1}{l}{1.456}&\multicolumn{1}{l}{1.457}&\multicolumn{1}{l}{1.445}&\multicolumn{1}{l}{0.76$\%$}&\multicolumn{1}{l}{0.82$\%$}\\
\multicolumn{1}{l}{\textbf{Movies and TV}}  &\multicolumn{1}{l}{2.803}&\multicolumn{1}{l}{2.334}&\multicolumn{1}{l}{1.721}&\multicolumn{1}{l}{1.723}&\multicolumn{1}{l}{1.682}&\multicolumn{1}{l}{2.27$\%$}&\multicolumn{1}{l}{2.38$\%$}\\
\multicolumn{1}{l}{\textbf{Home and Kitchen}} &\multicolumn{1}{l}{2.535}&\multicolumn{1}{l}{2.249}&\multicolumn{1}{l}{1.841}&\multicolumn{1}{l}{1.841}&\multicolumn{1}{l}{1.787}&\multicolumn{1}{l}{2.93$\%$}&\multicolumn{1}{l}{2.93$\%$}\\
\multicolumn{1}{l}{\textbf{CDs and Vinyl}}  &\multicolumn{1}{l}{2.508}&\multicolumn{1}{l}{2.083}&\multicolumn{1}{l}{1.746}&\multicolumn{1}{l}{1.740}&\multicolumn{1}{l}{1.523}&\multicolumn{1}{l}{12.77$\%$}&\multicolumn{1}{l}{12.47$\%$}\\
\multicolumn{1}{l}{\textbf{Cell Phones and Access.}} &\multicolumn{1}{l}{2.542}&\multicolumn{1}{l}{2.448}&\multicolumn{1}{l}{1.996}&\multicolumn{1}{l}{1.995}&\multicolumn{1}{l}{1.901}&\multicolumn{1}{l}{4.76$\%$}&\multicolumn{1}{l}{4.71$\%$}\\
\multicolumn{1}{l}{\textbf{Sports and Outdoors}}  &\multicolumn{1}{l}{2.138}&\multicolumn{1}{l}{2.096}&\multicolumn{1}{l}{1.422}&\multicolumn{1}{l}{1.422}&\multicolumn{1}{l}{1.349}&\multicolumn{1}{l}{5.13$\%$}&\multicolumn{1}{l}{5.13$\%$}\\
\multicolumn{1}{l}{\textbf{Kindle Store}}  &\multicolumn{1}{l}{2.516}&\multicolumn{1}{l}{2.145}&\multicolumn{1}{l}{1.814}&\multicolumn{1}{l}{1.813}&\multicolumn{1}{l}{1.581}&\multicolumn{1}{l}{14.12$\%$}&\multicolumn{1}{l}{14.12$\%$}\\
\multicolumn{1}{l}{\textbf{Health and Personal Care}}  &\multicolumn{1}{l}{2.392}&\multicolumn{1}{l}{2.259}&\multicolumn{1}{l}{1.670}&\multicolumn{1}{l}{1.689}&\multicolumn{1}{l}{1.666}&\multicolumn{1}{l}{0.24$\%$}&\multicolumn{1}{l}{1.36$\%$}\\
\multicolumn{1}{l}{\textbf{Apps for Android}} &\multicolumn{1}{l}{2.984}&\multicolumn{1}{l}{2.397}&\multicolumn{1}{l}{2.190}&\multicolumn{1}{l}{2.188}&\multicolumn{1}{l}{2.006}&\multicolumn{1}{l}{8.40$\%$}&\multicolumn{1}{l}{8.32$\%$}\\
\multicolumn{1}{l}{\textbf{Toys and Games}}  &\multicolumn{1}{l}{2.258}&\multicolumn{1}{l}{2.152}&\multicolumn{1}{l}{1.512}&\multicolumn{1}{l}{1.512}&\multicolumn{1}{l}{1.485}&\multicolumn{1}{l}{1.79$\%$}&\multicolumn{1}{l}{1.79$\%$}\\
\multicolumn{1}{l}{\textbf{Beauty}}  &\multicolumn{1}{l}{2.371}&\multicolumn{1}{l}{2.256}&\multicolumn{1}{l}{1.675}&\multicolumn{1}{l}{1.674}&\multicolumn{1}{l}{1.646}&\multicolumn{1}{l}{1.73$\%$}&\multicolumn{1}{l}{1.67$\%$}\\
\multicolumn{1}{l}{\textbf{Tools and Home Improve.}}  &\multicolumn{1}{l}{1.392}&\multicolumn{1}{l}{2.136}&\multicolumn{1}{l}{1.634}&\multicolumn{1}{l}{1.634}&\multicolumn{1}{l}{1.516}&\multicolumn{1}{l}{7.22$\%$}&\multicolumn{1}{l}{7.22$\%$}\\
\multicolumn{1}{l}{\textbf{Automotive}}  &\multicolumn{1}{l}{2.081}&\multicolumn{1}{l}{2.064}&\multicolumn{1}{l}{1.511}&\multicolumn{1}{l}{1.511}&\multicolumn{1}{l}{1.410}&\multicolumn{1}{l}{6.68$\%$}&\multicolumn{1}{l}{6.62$\%$}\\
\multicolumn{1}{l}{\textbf{Video Games}}  &\multicolumn{1}{l}{1.603}&\multicolumn{1}{l}{2.354}&\multicolumn{1}{l}{1.721}&\multicolumn{1}{l}{1.722}&\multicolumn{1}{l}{1.617}&\multicolumn{1}{l}{6.04$\%$}&\multicolumn{1}{l}{6.10$\%$}\\
\multicolumn{1}{l}{\textbf{Grocery and Gourmat}}  &\multicolumn{1}{l}{2.020}&\multicolumn{1}{l}{2.129}&\multicolumn{1}{l}{1.520}&\multicolumn{1}{l}{1.519}&\multicolumn{1}{l}{1.491}&\multicolumn{1}{l}{1.91$\%$}&\multicolumn{1}{l}{1.84$\%$}\\
\multicolumn{1}{l}{\textbf{Office Products}}  &\multicolumn{1}{l}{2.168}&\multicolumn{1}{l}{2.306}&\multicolumn{1}{l}{1.813}&\multicolumn{1}{l}{1.796}&\multicolumn{1}{l}{1.626}&\multicolumn{1}{l}{10.31$\%$}&\multicolumn{1}{l}{9.47$\%$}\\
\multicolumn{1}{l}{\textbf{Pet Supplies}}  &\multicolumn{1}{l}{2.480}&\multicolumn{1}{l}{2.265}&\multicolumn{1}{l}{1.815}&\multicolumn{1}{l}{1.814}&\multicolumn{1}{l}{1.765}&\multicolumn{1}{l}{2.81$\%$}&\multicolumn{1}{l}{2.70$\%$}\\
\multicolumn{1}{l}{\textbf{Patio, Lawn and Garden}}  &\multicolumn{1}{l}{1.970}&\multicolumn{1}{l}{2.235}&\multicolumn{1}{l}{1.780}&\multicolumn{1}{l}{1.781}&\multicolumn{1}{l}{1.738}&\multicolumn{1}{l}{2.36$\%$}&\multicolumn{1}{l}{-2.41$\%$}\\
\multicolumn{1}{l}{\textbf{Baby}}  &\multicolumn{1}{l}{1.934}&\multicolumn{1}{l}{2.194}&\multicolumn{1}{l}{1.717}&\multicolumn{1}{l}{1.715}&\multicolumn{1}{l}{1.693}&\multicolumn{1}{l}{1.40$\%$}&\multicolumn{1}{l}{1.28$\%$}\\
\multicolumn{1}{l}{\textbf{Digital Music}}  &\multicolumn{1}{l}{1.261}&\multicolumn{1}{l}{1.555}&\multicolumn{1}{l}{1.030}&\multicolumn{1}{l}{1.030}&\multicolumn{1}{l}{1.052}&\multicolumn{1}{l}{-2.14$\%$}&\multicolumn{1}{l}{-2.14$\%$}\\
\multicolumn{1}{l}{\textbf{Amazon Instant Video}}  &\multicolumn{1}{l}{2.828}&\multicolumn{1}{l}{1.985}&\multicolumn{1}{l}{1.278}&\multicolumn{1}{l}{1.277}&\multicolumn{1}{l}{1.508}&\multicolumn{1}{l}{-18.09$\%$}&\multicolumn{1}{l}{-18.00$\%$}\\
\multicolumn{1}{l}{\textbf{Musical Instruments}}  &\multicolumn{1}{l}{1.735}&\multicolumn{1}{l}{1.921}&\multicolumn{1}{l}{1.165}&\multicolumn{1}{l}{1.165}&\multicolumn{1}{l}{1.208}&\multicolumn{1}{l}{-3.69$\%$}&\multicolumn{1}{l}{-3.69$\%$}\\
\hline
\multicolumn{1}{l}{\textbf{Average MSE K = 5}}  &\multicolumn{1}{l}{2.248}&\multicolumn{1}{l}{2.122}&\multicolumn{1}{l}{1.643}&\multicolumn{1}{l}{1.644}&\multicolumn{1}{l}{1.572}\\
\hline
\multicolumn{1}{l}{\textbf{Average MSE K = 10}}  &\multicolumn{1}{l}{2.248}&\multicolumn{1}{l}{2.122}&\multicolumn{1}{l}{1.637}&\multicolumn{1}{l}{1.637}&\multicolumn{1}{l}{1.579}\\
\hline
\end{tabular}
\caption[Recommender Systems Prediction Results]{Prediction results in terms of MSE with K = 5 number topics. Imp.$_{[5]/[3]}$ reports the percentage improvement of the LDA-LFM model compared to the LFM model, in terms of prediction accuracy. Imp.$_{[5]/[4]}$ shows the percentage improvement of the LDA-LFM model compared to the LDAFirst model. The average MSE per model is also reported for the K = 10 case.}
\label{ress1}
\end{table*}
\begin{table*}[h]
\ra{1.18}
\footnotesize\centering
\begin{tabular}{|p{11cm}|p{7cm}|p{7cm}|p{7cm}|p{7cm}|p{7cm}|}
\hline
\multicolumn{1}{l}{\textbf{Dataset}} &\multicolumn{1}{l}{\textbf{K$^*$= 0}}&\multicolumn{1}{l}{\textbf{K$^*$= 1}}&\multicolumn{1}{l}{\textbf{K$^*$= 2}}&\multicolumn{1}{l}{\textbf{K$^*$= 3}}&\multicolumn{1}{l}{\textbf{K$^*$= 4}}\\
\hline
\multicolumn{1}{l}{\textbf{Electronics}} &\multicolumn{1}{l}{1.78019}&\multicolumn{1}{l}{\textbf{1.78001}}&\multicolumn{1}{l}{\textbf{1.77651}}&\multicolumn{1}{l}{\textbf{1.77901}}&\multicolumn{1}{l}{1.78199}\\
\multicolumn{1}{l}{\textbf{Clothing, Shoes and Jewel.}} &\multicolumn{1}{l}{1.44511}&\multicolumn{1}{l}{\textbf{1.44510}}&\multicolumn{1}{l}{\textbf{1.44515}}&\multicolumn{1}{l}{\textbf{1.44510}}&\multicolumn{1}{l}{1.44581}\\
\multicolumn{1}{l}{\textbf{Movies and TV}}  &\multicolumn{1}{l}{1.68230}&\multicolumn{1}{l}{\textbf{1.68204}}&\multicolumn{1}{l}{1.69210}&\multicolumn{1}{l}{1.69199}&\multicolumn{1}{l}{\textbf{1.68221}}\\
\multicolumn{1}{l}{\textbf{Home and Kitchen}} &\multicolumn{1}{l}{1.78696}&\multicolumn{1}{l}{\textbf{1.78694}}&\multicolumn{1}{l}{\textbf{1.78690}}&\multicolumn{1}{l}{1.78710}&\multicolumn{1}{l}{\textbf{1.78691}}\\
\multicolumn{1}{l}{\textbf{CDs and Vinyl}}  &\multicolumn{1}{l}{1.52294}&\multicolumn{1}{l}{\textbf{1.52293}}&\multicolumn{1}{l}{\textbf{\textbf{1.52278}}}&\multicolumn{1}{l}{\textbf{1.52280}}&\multicolumn{1}{l}{\textbf{1.52270}}\\
\multicolumn{1}{l}{\textbf{Cell Phones and Access.}} &\multicolumn{1}{l}{1.90113}&\multicolumn{1}{l}{1.90553}&\multicolumn{1}{l}{1.90540}&\multicolumn{1}{l}{\textbf{1.90112}}&\multicolumn{1}{l}{ \textbf{1.90154}}\\
\multicolumn{1}{l}{\textbf{Sports and Outdoors}} &\multicolumn{1}{l}{1.34960}&\multicolumn{1}{l}{\textbf{1.34959}}&\multicolumn{1}{l}{\textbf{1.34950}}&\multicolumn{1}{l}{\textbf{1.39512}}&\multicolumn{1}{l}{\textbf{1.39516}}\\
\multicolumn{1}{l}{\textbf{Kindle Store}}  &\multicolumn{1}{l}{1.58104}&\multicolumn{1}{l}{1.58118}&\multicolumn{1}{l}{1.58143}&\multicolumn{1}{l}{\textbf{1.58100}}&\multicolumn{1}{l}{\textbf{1.58011}}\\
\multicolumn{1}{l}{\textbf{Health and Personal Care}}&\multicolumn{1}{l}{1.66564}&\multicolumn{1}{l}{\textbf{1.66558}}&\multicolumn{1}{l}{\textbf{1.66557}}&\multicolumn{1}{l}{\textbf{1.66555}} &\multicolumn{1}{l}{\textbf{1.66553}}\\
\multicolumn{1}{l}{\textbf{Apps for Android}} &\multicolumn{1}{l}{2.00655}&\multicolumn{1}{l}{\textbf{2.00617}}&\multicolumn{1}{l}{\textbf{2.00524}}&\multicolumn{1}{l}{\textbf{2.00644}} &\multicolumn{1}{l}{2.00696}\\
\multicolumn{1}{l}{\textbf{Toys and Games}} &\multicolumn{1}{l}{1.48472}&\multicolumn{1}{l}{\textbf{1.48455}}&\multicolumn{1}{l}{\textbf{1.48458}}&\multicolumn{1}{l}{\textbf{1.48442}}&\multicolumn{1}{l}{\textbf{1.48446}}\\
\multicolumn{1}{l}{\textbf{Beauty}}  &\multicolumn{1}{l}{1.64578}&\multicolumn{1}{l}{1.64584}&\multicolumn{1}{l}{1.64580}&\multicolumn{1}{l}{1.64578}&\multicolumn{1}{l}{1.64583}\\
\multicolumn{1}{l}{\textbf{Tools and Home improve.}}  &\multicolumn{1}{l}{1.51640}&\multicolumn{1}{l}{1.51644}&\multicolumn{1}{l}{\textbf{1.51637}}&\multicolumn{1}{l}{1.51658}&\multicolumn{1}{l}{1.51656}\\
\multicolumn{1}{l}{\textbf{Automotive}} &\multicolumn{1}{l}{1.40996}&\multicolumn{1}{l}{\textbf{1.40983}}&\multicolumn{1}{l}{\textbf{1.40977}}&\multicolumn{1}{l}{\textbf{1.40973}}&\multicolumn{1}{l}{\textbf{1.40984}}\\
\multicolumn{1}{l}{\textbf{Video Games}} &\multicolumn{1}{l}{1.61691}&\multicolumn{1}{l}{1.61699}&\multicolumn{1}{l}{\textbf{1.60912}}&\multicolumn{1}{l}{\textbf{1.60902}}&\multicolumn{1}{l}{\textbf{1.60936}}\\
\multicolumn{1}{l}{\textbf{Grocery and Gourmet}} &\multicolumn{1}{l}{1.49049}&\multicolumn{1}{l}{1.49057}&\multicolumn{1}{l}{\textbf{1.49046}}&\multicolumn{1}{l}{\textbf{1.49045}}&\multicolumn{1}{l}{1.49084}\\
\multicolumn{1}{l}{\textbf{Office Products}} &\multicolumn{1}{l}{1.62635}&\multicolumn{1}{l}{1.62666}&\multicolumn{1}{l}{1.62673}&\multicolumn{1}{l}{1.62674}&\multicolumn{1}{l}{1.62668}\\
\multicolumn{1}{l}{\textbf{Pet Supplies}}  &\multicolumn{1}{l}{1.76503}&\multicolumn{1}{l}{\textbf{1.76502}}&\multicolumn{1}{l}{1.76501}&\multicolumn{1}{l}{1.76500}&\multicolumn{1}{l}{\textbf{1.76515}}\\
\multicolumn{1}{l}{\textbf{Patio, Lawn and Garden}} &\multicolumn{1}{l}{1.73829}&\multicolumn{1}{l}{\textbf{1.73810}}&\multicolumn{1}{l}{\textbf{1.73812}}&\multicolumn{1}{l}{\textbf{1.73811}}&\multicolumn{1}{l}{\textbf{1.73809}}\\
\multicolumn{1}{l}{\textbf{Baby}}  &\multicolumn{1}{l}{1.69345}&\multicolumn{1}{l}{\textbf{1.69332}}&\multicolumn{1}{l}{1.69358}&\multicolumn{1}{l}{\textbf{1.69272}}&\multicolumn{1}{l}{1.69372}\\
\multicolumn{1}{l}{\textbf{Digital Music}} &\multicolumn{1}{l}{1.05179}&\multicolumn{1}{l}{\textbf{1.05165}}&\multicolumn{1}{l}{\textbf{1.05175}}&\multicolumn{1}{l}{1.05185}&\multicolumn{1}{l}{\textbf{1.05170}}\\
\multicolumn{1}{l}{\textbf{Amazon Instant Video}} &\multicolumn{1}{l}{1.50755}&\multicolumn{1}{l}{1.50852}&\multicolumn{1}{l}{1.50949}&\multicolumn{1}{l}{1.50835}&\multicolumn{1}{l}{1.50885}\\
\multicolumn{1}{l}{\textbf{Musical Instruments}}  &\multicolumn{1}{l}{1.20791}&\multicolumn{1}{l}{\textbf{1.19803}}&\multicolumn{1}{l}{\textbf{1.19825}}&\multicolumn{1}{l}{\textbf{1.19818}}&\multicolumn{1}{l}{\textbf{1.19817}}\\
\hline
\multicolumn{1}{l}{\textbf{Number of Times Beneficial}}  &\multicolumn{1}{l}{-}&\multicolumn{1}{l}{15}&\multicolumn{1}{l}{15}&\multicolumn{1}{l}{15}&\multicolumn{1}{l}{14}\\
\hline
\end{tabular}
\caption[Recommender Systems Prediction Results with Extra Features]{Prediction results of the LDA-LFM model in terms of the MSE with K = 5 number topics. K$^*$ defines the number of extra factors added to the LDA-LFM model. K$^*$ = 0 represents the case when no extra feature has been added to the model, results corresponding to the Table \ref{ress1}. K$^*$ = 1, K$^*$ = 2, K$^*$ = 3, and K$^*$ = 4 correspond to the LDA-LFM model predictions with 1, 2, 3, and 4 extra features, respectively.}\label{ress4}
\end{table*}

Table \ref{ress4} presents the prediction results in terms of MSE, per product category with the number of topics equal to 5 and extra added features. We observe that for almost all datasets there is at least one LFM-LDA model with extra feature(s) with higher prediction accuracy (lower MSE value compared to the corresponding MSE value in the K$^*$ = 0 model, i.e., without extra features). Moreover, for the  datasets \textit{Musical Instruments}, \textit{Patio, Lawn and Garden}, \textit{Automotive}, \textit{Toys and Games}, \textit{Health and Personal Care}, \textit{Sports and Outdoors}, \textit{CDs and Vinyls}, \textit{Home and Kitchen} and \textit{Movies and TV} all four models with different number of extra features are performing better compared to the model without extra added features. However, there are few datasets (\textit{Amazon Instant Video}, \textit{Office Products}, and \textit{Beauty}) for which it holds that adding extra features to the LDA-LFM model either does not change or worsens the performance of the model. We observe that all those datasets, for which adding extra features is not efficient, are either very small or medium size datasets in the set of all 23 Amazon datasets used in this study. The last row of Table \ref{ress4} presents the number of cases in which adding a particular amount of extra features leads to an increase in the prediction accuracy of the model. We observe that in all four cases (adding 1, 2, 3, and 4 extra features), the number of datasets with better performance are close to each other. More specifically about 15 out of 23 datasets (from which around 9 cases corresponds to a medium or a large dataset), adding extra features results in more accurate recommendations.

\section{Conclusion}
\label{section: conclusions}

Taking into account the current limitations in the existing literature, in this paper, we utilize the LFM model by using both ratings and textual reviews of customers, such that it is applicable on both small and large datasets and also allows adding user- and item-specific data to it. We have shown how one can combine review-based LDA for topic modeling, with rating-based LFM for rating predictions. From the prediction results where we compared the prediction accuracy of the proposed LDA-LFM model to the prediction accuracy's of various baseline models applied on various datasets. We found that adding textual reviews to the recommender system leads to an increased prediction accuracy, which is especially true for medium and large datasets. Then we introduced an approach of adding extra latent features to the user-item rating matrix of the proposed LDA-LFM model, representing the user- and item-specific features not present in the review data. We found that for the majority of datasets (15 out of 23 datasets) it holds that, adding extra features to the proposed recommender system increases the quality of its recommendations, resulting in lower MSE, thus higher prediction accuracy. This indicates that again the improvements are better visible in medium and large datasets.

As future work we would like to investigate whether using sentiment analysis improves the quality of recommendations. Correspondingly, one can extend our model
so that it combines the rating modeling, topic extraction, and sentiment analysis techniques for making recommendations. For instance, sentiment analysis can be used to classify whether a review is negative or positive. \cite{ziani2017} proposed a recommender system combining rating based CF system with sentiment analysis for making recommendations. \cite{Lin2009} introduced the Joint Sentiment/Topic (JST) model which combines the topic modeling method LDA with sentiment analysis in order to detect a topic and a sentiment from
text simultaneously. For future work, we aim to combine the JST model with our
model for potentially
better recommendations. In this way, we can exploit topics that carry sentiment and are possibly better proxies for
ratings.

\bibliographystyle{ACM-Reference-Format}
\bibliography{references}
\end{document}